\documentclass[aps,prapplied, twocolumn, groupedaddress, showkeys, longbibliography]{revtex4-2}

\usepackage{graphics}
\usepackage{physics}
\usepackage{color}
\usepackage{verbatim}
\usepackage{subfiles}
\usepackage{amsmath}

\RequirePackage{siunitx}
\usepackage[hidelinks]{hyperref}
\usepackage[
    type={CC},
    modifier={by},
    version={4.0},
]{doclicense}
\usepackage{orcidlink}

\begin{document}
\doclicenseThis

\title{Dipolar coupled core-shell perpendicular shape anisotropy MTJ with enhanced write speed and reduced cross-talk}

\author{N. Ca\c{c}oilo\,\orcidlink{0000-0001-8847-2034}}
\author{L. D. Buda-Prejbeanu\,\orcidlink{0000-0002-6105-151X}}
\author{B. Dieny\,\orcidlink{0000-0002-0575-5301}}
\author{O. Fruchart\,\orcidlink{0000-0001-7717-5229}}
\author{I. L. Prejbeanu\,\orcidlink{0000-0001-6577-032X}}
\email[lucian.prejbeanu@cea.fr]{}
\affiliation{Univ. Grenoble Alpes, CEA, CNRS, Grenoble-INP, SPINTEC, 38000 Grenoble, France}

\date{\today}

\begin{abstract}
The concept of Perpendicular Shape-Anisotropy Spin-Transfer-Torque Magnetic Random-Access Memory tackles the downsize scalability limit of conventional ultrathin magnetic tunnel junctions (MTJ) below sub-$\SI{20}{nm}$ technological nodes. This concept uses a thicker storage layer with a vertical aspect ratio, enhancing the thermal stability factor $\Delta$ thanks to the favorable contribution of the shape anisotropy. However, the increased aspect ratio comes with an increase in switching time under applied voltage and the cross-over to non-uniform reversal mechanism at higher aspect ratio, limiting the gain in scalability. Additionally, the larger volume of the magnetic cell significantly increases the stray field acting on the neighboring devices compared to thin MTJs. In this work, we propose the use of a dipolar-coupled core-shell system as a storage layer. This improves both bottlenecks, as predicted by micromagnetic simulations for  magnetisation reversal, and a macrospin model to estimate the stray field in a dense array.
\end{abstract}


\maketitle

\section{Introduction}

Spin-Transfer-Torque Magnetic Random-Access Memory (STT-MRAM) is one of the most promising emerging nonvolatile memory technologies. It offers nonvolatility with very large write endurance, high speed, moderate write power, and zero off consumption \cite{dieny_opportunities_2020}. The retention time is related to the stability factor $\Delta$ of the device, defined as the ratio between the energy barrier ($E_\mathrm{B}$) of the storage layer and the thermal energy ($\textrm{k}_\mathrm{B}\mathrm{T}$, where $\textrm{k}_\mathrm{B}$ is Boltzmann constant and $T$ the operating temperature). The energy barrier for usual perpendicular magnetic tunnel junctions, made of a stack of ultra-thin layers with perpendicular magnetisation (p-MTJ), can be expressed as, assuming coherent reversal:
\begin{equation}
    E_\mathrm{B} = AL\left[\frac{k_\mathrm{s}}{L} + \frac{1}{2} \mu _0 M_\mathrm{s}^2 \left( N_{xx} - N_{zz}\right) + K_\mathrm{u} \right]\;,
    \label{eq:EnergyBarrier}
\end{equation}
where $A$ is the device area of the magnetic storage layer, $L$ its height, $k_s$ its surface anisotropy per unit area, $\mu _0$ the vacuum permeability, $M_\mathrm{s}$ the spontaneous magnetisation, $N_{ii}$ its demagnetising factors along the plane ($xx$) and out-of-plane ($zz$), and $K_\mathrm{u}$ the bulk uniaxial anisotropy per unit volume. The latter is usually small in conventional p-MTJs, leaving a competition between surface and shape anisotropy. For conventional p-MTJ, the lateral dimension of the storage layer is much larger than its height, resulting in a negative and large contribution of the shape anisotropy. This lowers the stability of the out-of-plane state, which is only promoted by surface anisotropy. When the p-MTJ lateral dimension is reduced, the leading variation in Eq. \ref{eq:EnergyBarrier} comes through the decrease of $A$, which decreases the stability. In practice, this limits the scalability of the p-MTJ to diameters of around \SI{20}{nm} \cite{sato_magnetic_2017}. A proposal to pushback this limit is to provide an additional source of total anisotropy energy and thus energy barrier, via the use of a thicker storage layer. Inspecting Eq.(\ref{eq:EnergyBarrier}), we see that this works out via two aspects: (1)~Decrease, in absolute value, the negative and thus destabilising impact of the in-plane shape effect per unit volume~(the right part, in brackets), or even make it positive for $L/D >  0.89$~(for a disk shape, with $D$ the cell diameter and thus $A=\pi D^2/4$) (2)~Increase the volume prefactor $AL$, which has a beneficial impact to enhance the energy barrier in the case of perpendicular shape anisotropy, i.e. for $L/D >  0.89$. This concept has been proven to be effective to extend the range of stability down to sub-$\SI{10}{nm}$ diameter \cite{Perrissin, Watanabe}. However, this comes at the expense of the need for a larger writing voltage for STT switching, and longer switching time, both being a handicap for applicability. 

In this manuscript, we propose an alternative approach to address the writing drawbacks associated with the thick PSA pillars required at small diameter, introducing a tubular magnetic shell around the storage layer. Making use of the strong dipolar coupling existing in this system thanks to its high vertical aspect ratio, it is possible to promote a core-shell synthetic ferrimagnetic state, with antiparallel storage layer (from now on described as magnetic \textit{core}) and shell. The gain introduced by this design is twofold. First, antiparallel coupling enhances stability. This allows a decrease of the total height of the core-shell system to reach a given stability, reducing the volume to switch and delaying the onset of the domain-wall-based reversal mechanism, thus leading to a lower voltage and faster reversal. Second, the stray field emanating from the composite cell is sharply reduced, enabling a denser array than would be possible for single-pillar PSA MTJs. In the following we first review and refine the evaluation of cell stability~(section~\ref{sec-stability}), we then present the core-shell concept and the resulting gain in writing~(section~\ref{sec-coreShellConcept}), and finally describe the additional gain via the reduciton of stray field and thus cross-talk~(section~\ref{sec-crossTalk}).

\section{Capping in scalability of a single PSA storage layer}
\label{sec-stability}

To understand the increased stability quantitatively, it is necessary to calculate the energy barrier of the system with a focus on the magnetostatic energy (related to the shape anisotropy). To calculate this term, we follow the approach used in \cite{Beleggia2005, TandonPt1}, where a Fourier-space formalism based on the shape amplitude $\mathcal{D}(\mathbf{k})$ (in the k-space) of the object is used to compute the demagnetising factors. For a cylinder, the shape amplitude is given by (in cylindrical coordinates):
\begin{equation}
    \mathcal{D}(k_\perp, k_z) = \frac{4\pi R}{k_\perp k_z} \mathcal{J}_1 (k_\perp R) \sin\left(\frac{L}{2}k_z\right)\;,
    \label{eq:ShapeAmplitude}
\end{equation}
with $k_\perp = \sqrt{k_x^2 + k_y ^2}$ related to direction perpendicular to the cylinder axis, $R$ the radius of the magnetic cylinder and $\mathcal{J}_1(x)$ the Bessel function of the first kind. It is then possible to calculate the demagnetising factor along the direction along the axis of the cylinder:
\begin{equation}
    \mathcal{N}_{zz} = \frac{1}{8\pi^3V}\int \frac{\mathrm{d}^3 \textbf{k}}{\textbf{k}^2} | \mathcal{D}(\textbf{k})|^2k_z^2.
\end{equation}

After a series of integrations and transformations, it is possible to achieve a simpler notation like in \cite{Beleggia2005}

\begin{equation}
    \mathcal{N}_{zz}^{\mathrm{cylinder}} = 1 + \frac{4}{3\pi\tau} - F_{0}\left(-\frac{1}{\tau^2}\right),
\end{equation}
where $F_{0}$ is the hyper-geometric function, shown here for convenience: 
\begin{equation*}
    F_0(x) = {_{2}F^1}\left( -\frac{1}{2}, \frac{1}{2}, 2, x\right). 
\end{equation*}

\begin{figure}
    \centering
    \includegraphics{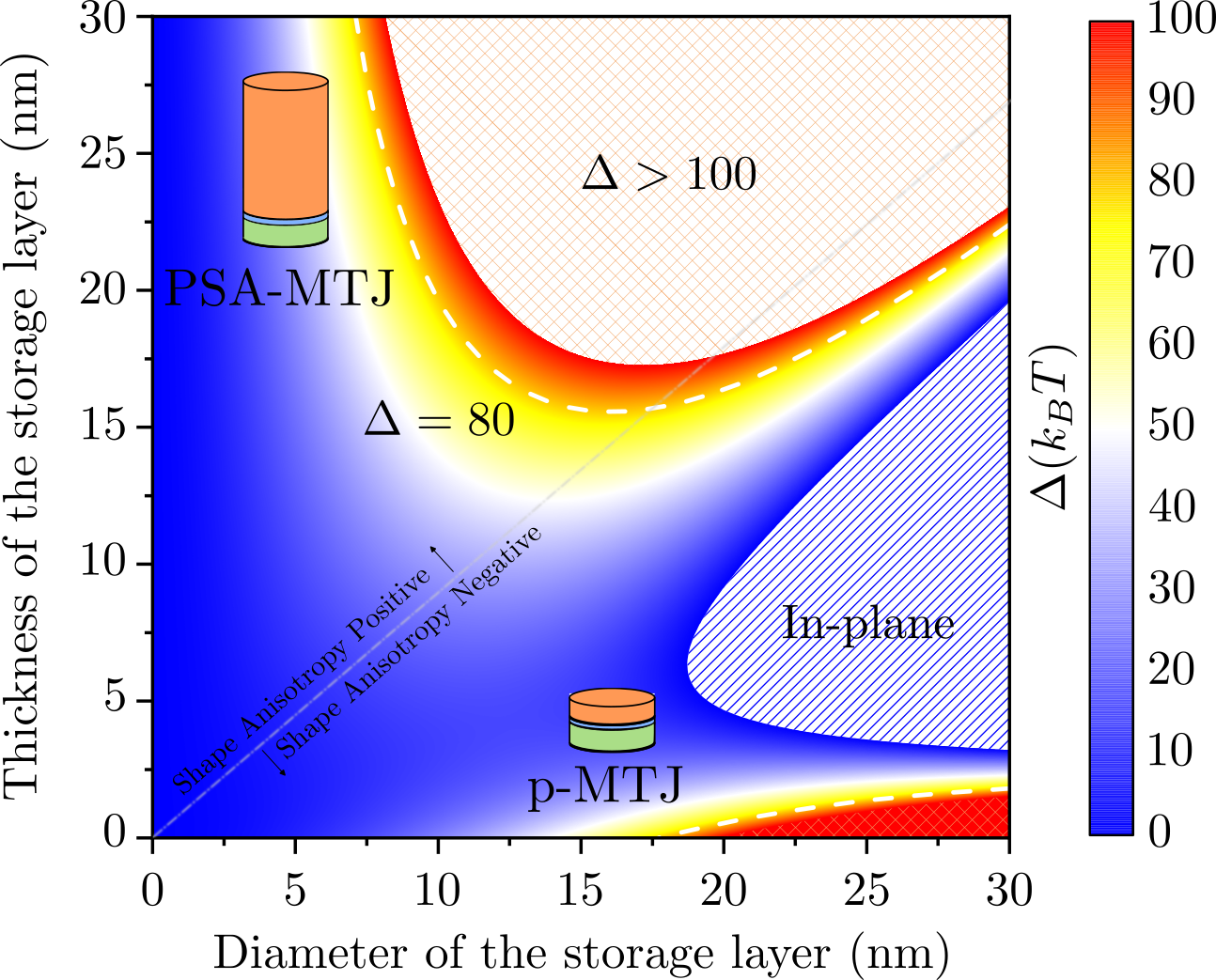}
    \caption{Stability factor for different dimensions of the storage layer with a spontaneous magnetisation of $\SI{1}{MA/m}$ for a single MgO interface with a surface anisotropy of $\SI{1.4}{mJ/m^2}$. The dashed line (diagonal) shows the limit between regions with positive and negative values of the magnetostatic energy.}
    \label{fig:figure 1}
\end{figure}

The two identical transverse demagnetising coefficients can be calculated as $(1-\mathcal{N}_{zz}^{\mathrm{cylinder}})/2$, and the magnetostatic anisotropy energy coefficient scales with $(1-3\mathcal{N}_{zz}^{\mathrm{cylinder}})/2$. Figure \ref{fig:figure 1} shows a stability diagram depending on the geometry of the storage layer (height and diameter) for a storage layer with the usual FeCoB material with spontaneous magnetisation of $\SI{1}{MA/m}$ and a surface anisotropy of $\SI{1.4}{mJ/m^2}$, a typical value in MgO$|$FeCoB systems. From it, relevant information can be extracted. This diagram illustrates the limitation of the usual p-MTJ at small nodes as the stability is reduced with the diameter, going below the required levels at around \SI{15}{nm} diameter (typical for a 10 year retention time values around 60-80 $\textrm{k}_\textrm{B}\textrm{T}$ are used). As the height is increased there is first a reduction in perpendicular stability, as the magnetostatic energy is still favouring the in-plane orientation, but now its influence is enhanced by the larger volume of the storage layer. This can induce an in-plane alignment of the magnetisation, as seen for $D > \SI{18}{nm}$. However, if the thickness is further increased there is a stability increase, coming from the progressive reduction of the difference in axis versus transverse demegnetising factors or even its cheage in sign, so that perpendicular magnetisation is restored as the easy axis. The height necessary to restore perpendicular orientation reduces with reducing diameter (this orientation is aspect-ratio dependent). Below $D=\SI{18}{nm}$ the easy axis always remains perpendicular. For a cylindrical pillar with a reasonable high surface anisotropy value (as in the example of $\SI{1.4}{mJ/m^{2}}$) it is possible to maintain a perpendicular orientation with a large stability down to sub-\SI{5}{nm} diameter. Although at first glance an effective path for high density, altogether with an exceptional tolerance for high temperature applications (since the thermal dependence of the magnetisation is less steep than that of the surface anisotropy \cite{almeida_quantitative_2022, StevenIMW, Steven2020}), there are still several obstacles that need to be addressed to push this technology forward. 

\begin{figure}
    \centering
    \includegraphics{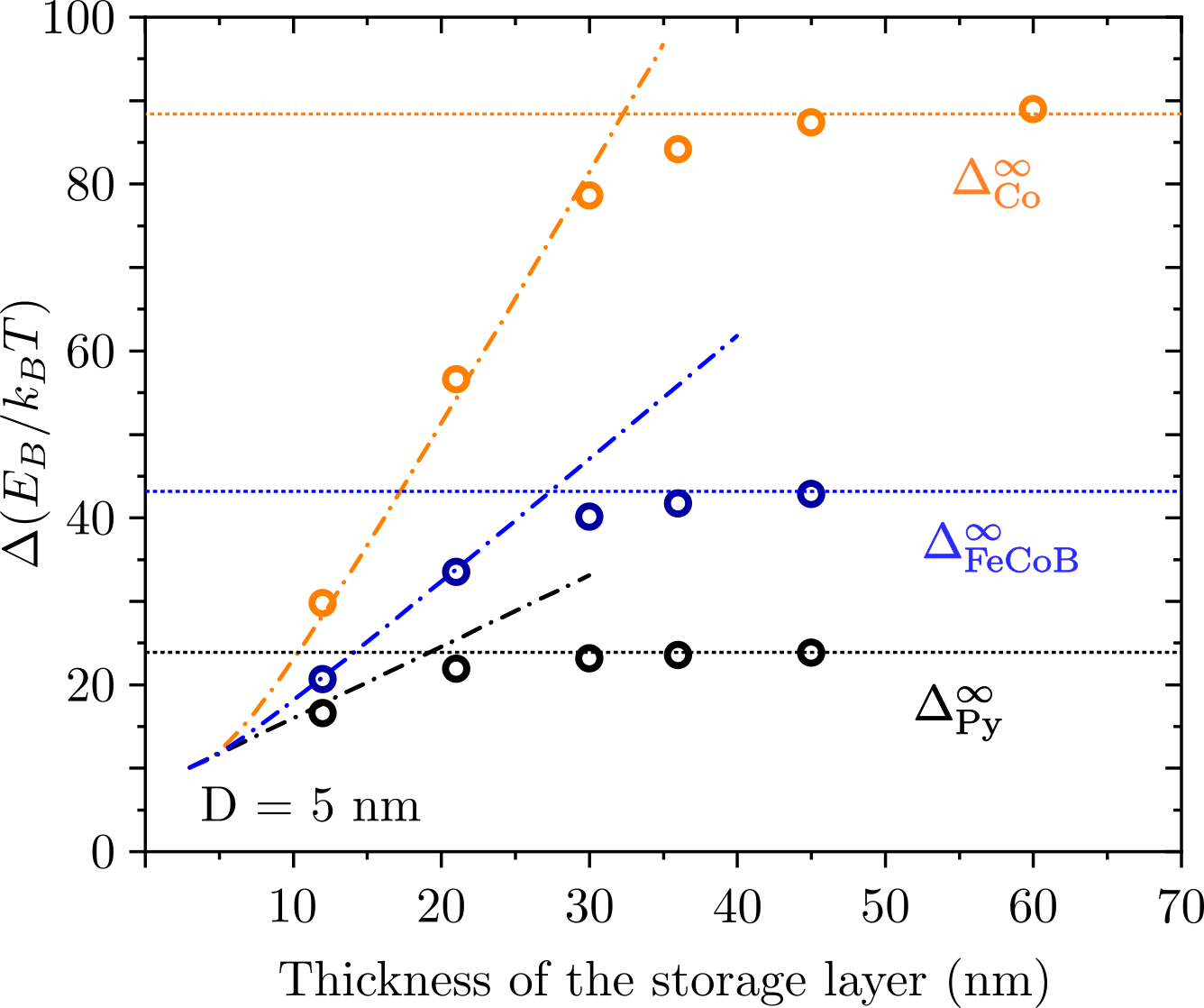}
    \caption{Stability factor for different PSA storage layers for a \SI{5}{nm} diameter storage layer with FeCoB double interface ($\SI{1.4}{mJ/m^2}$ and $\SI{1.0}{mJ/m^2}$). Predictions based on macrospin are shown with dashed-dotted lines, while the capping of stability related to domain wall nucleation-propagation is shown with the horizontal dotted lines. The open disks stand for micromagnetic simulations of minimum energy path.}
    \label{fig:figure 2}
\end{figure}

One of the drawbacks of a high aspect ratio PSA pillar is to achieve fast and efficient switching with spin transfer torque, as it is largely an interfacial effect in MTJs \cite{Nuno}. Moreover, there is a capping in stability, due to the formation of a transient domain-wall along the height of the storage layer during the magnetisation reversal. As already discussed \cite{Perrissin}, this capping of stability can be determined considering a tail-to-tail domain wall formed along the thickness of the magnetic body, which is dependent on the exchange stiffness $A_\mathrm{ex}$ and the spontaneous magnetisation $M_\mathrm{s}$ of the magnetic cylinder:
\begin{equation}
    \Delta^{\mathrm{DW}} = \frac{\mu_0M_\textrm{s}^2}{2k_\textrm{B}T}\frac{\pi D^2}{4}\left(\frac{D}{2}+L_{\mathrm{DW}} + \frac{2L_{\mathrm{DW}}^2}{D + 2L_\mathrm{DW}}\right)\;,
\end{equation}
where $L_\mathrm{DW}$ is the width of the domain wall, scaling with $\sqrt{\frac{4A_\textrm{ex}}{\mu_0 M_\textrm{s}^2}}$. As the height at which the capping sets in is related to the domain wall width, a workaround would be to reduce the total height by increasing the surface anisotropy, for example, through an additional MgO interface at the top of the storage layer (as in common double interface p-MTJ). Experimental efforts have been reported in this direction \cite{Jinnai2020, Jinnai2021}. However, this approach provides moderate gain at lower diameters, as the contribution of the surface anisotropy is less significant. This is observed in Fig. \ref{fig:figure 2}, where the stability capping is shown for different thick layers (Co, FeCoB and Py), with double FeCoB interfaces, with different surface anisotropies of $\SI{1.4}{mJ/m^2}$ and $\SI{1.0}{mJ/m^2}$. On this graph, we also see that the lower the magnetisation, the larger the height needs to be to achieve 
a given stability, as for these diameters the main anisotropy source is magnetostatic.

\section{Core-Shell concept for storage layer}
\label{sec-coreShellConcept}

To understand the increase in scalability provided by the magnetic shell, it is helpfull to first calculate its stability as an isolated element. Thus, we need to calculate its demagnetising factors, since the shape anisotropy is the only source of magnetic anisotropy. To do so, we use the shape amplitude of equation \ref{eq:ShapeAmplitude}, with outer radius $R_2$ and inner radius $R_1$ \cite{beleggia_phase_2006, BeleggiaShell2009}:

\begin{multline}
        \mathcal{D}(k_\perp, k_z) = \frac{4\pi}{k_\perp k_z} \big[(R_2\mathcal{J}_1 (k_\perp R_2) - \\ R_1\mathcal{J}_1 (k_\perp R_1)\big]\sin{(\frac{L}{2}k_z)}\;.
        \label{eq:ShapeAmplitude-shell}
\end{multline}
Following the same reasoning as before, and performing the transformations $q = k_\perp R_2$, $\sigma = R_1 / R_2$, and $\tau = L/(2R_2)$ we obtain the integral:

\begin{multline}
    \mathcal{N}_{zz}^{\textrm{shell}}(\sigma, \tau) = \frac{1}{\tau(1-\sigma^2)} \int^{+ \infty }_{0} \frac{\mathrm{d}q}{q^2} (1 - e^{-2q\tau}) \\ \big[(\mathcal{J}_1(q) - \sigma\mathcal{J}_1(\sigma q)\big]^2\;.
\end{multline}

The latter can be expanded to be written with the demagnetising coefficients of full cylinders ($\sigma=0$):
\begin{multline}
    \mathcal{N}_{zz}^{\mathrm{shell}}(\sigma, \tau) = \frac{1}{1-\sigma^2} \Bigg[\mathcal{N}_{zz}^\textrm{cylinder}\left(\tau \right) + \sigma^2\mathcal{N}_{zz}^\textrm{cylinder}\left(\frac{\tau}{\sigma}\right) - \\ \frac{\sigma^2}{\tau}F_0(\sigma^2) + \frac{2\sigma}{\tau}A_{1,1}^{-1}(2\tau, 1, \sigma)\Bigg],
    \label{eq:Shell-demag}
\end{multline}
with:
\begin{multline}
    A_{1,1}^{-1}(2\tau, 1, \sigma) = \int^{+ \infty }_{0}  q^{-2}e^{2q\tau}\mathcal{J}_1(q)\mathcal{J}_1(\sigma q) \mathrm{d}q\;,
    \label{eq:A11}
\end{multline}
which can be solved making use of a combination of elliptical integrals (see, for reference, equation 10 and table 1 of \cite{BeleggiaShell2009}). 

\begin{figure}
    \centering
    \includegraphics{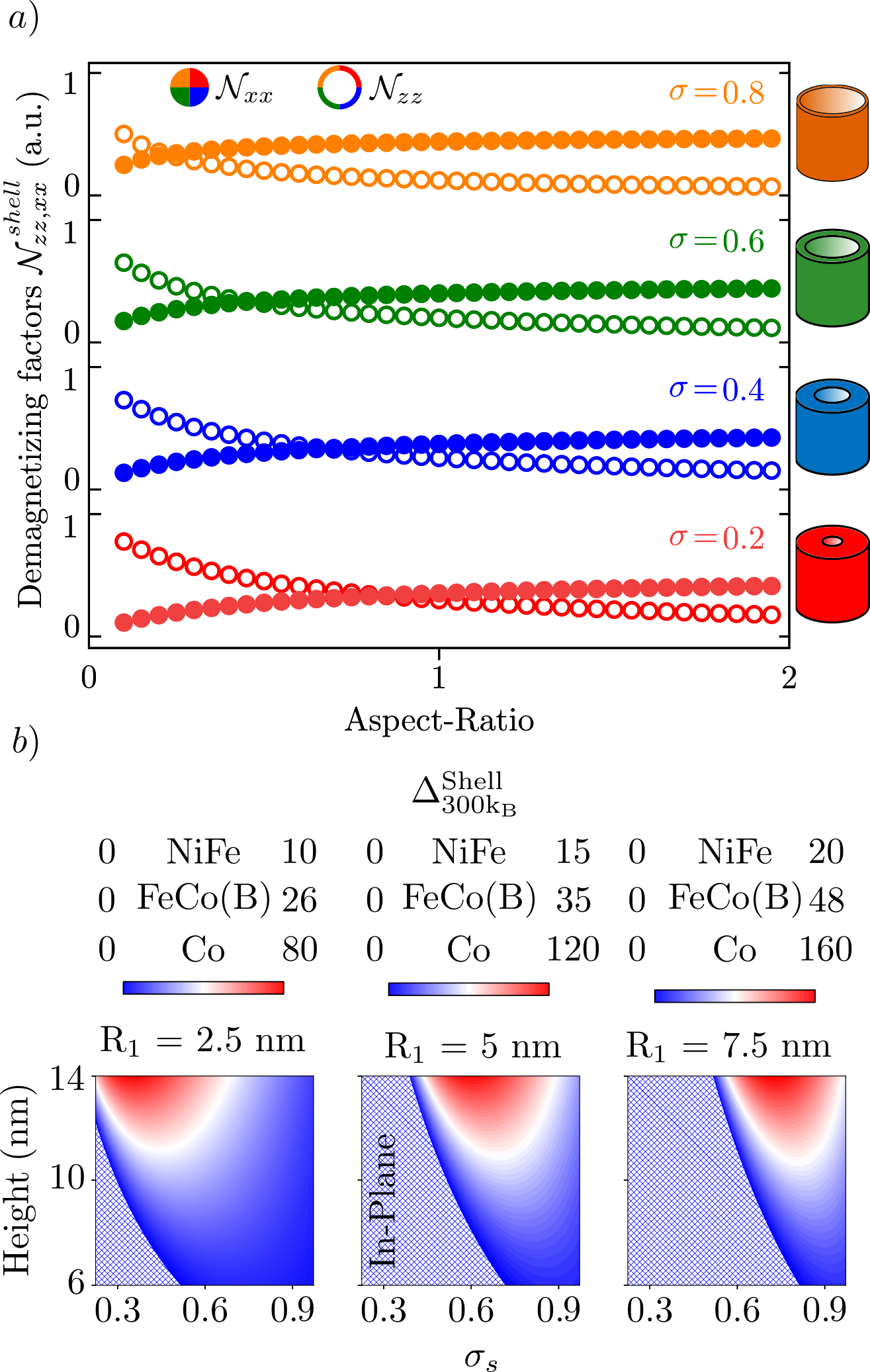}
    \caption{$a)$ Demagnetising factors $\mathcal{N}_{xx}$ (filled disks) and $\mathcal{N}_{zz}$ (open disks)  for the magnetic shell as a function of the aspect-ratio $\tau= L/(2R_2)$ for different  for different values of $\sigma$ and $b)$ Stability factor of the shell as a function of the shell thickness ratio $\sigma$ and aspect-ratio $\tau^\textrm{shell}$ for three different inner radius $R_1$ values for the situation of a Ni shell $\SI{0.489}{MA/m}$, NiFe shell $\SI{0.756}{MA/m}$ shell and Co shell $\SI{1.446}{MA/m}$. Patterned region showns an in-plane orientation of the magnetisation.}
    \label{fig:figure 3}
\end{figure}

Using equation \ref{eq:Shell-demag} it is possible to calculate the preferred orientation of the magnetisation in the magnetic shell, as a function of the aspect ratio $\tau$ and $\sigma$. Figure \ref{fig:figure 3} illustrates this versus aspect ratio, for several values of $\sigma$. There are two important features to be extracted: The crossing point between the  $\mathcal{N}_{xx}^{\textrm{shell}}$ and $\mathcal{N}_{zz}^{\textrm{shell}}$ curves highlights the crossover between in-plane versus perpendicular easy directions. This occurs at smaller aspect-ratios as the value of $\sigma$ gets larger, as the shell resembles a thin rolled sheet. In the opposite situation of small $\sigma$, the aspect-ratio for crossover is similar to that of a full cylinder. Above the crossover, the difference between the two demagnetising factors keeps increasing as the aspect ratio increases, liable to provide a larger stability to the magnetic shell. 

This stability, product of the volume of the shell and the difference of its demagnetising coefficients shown in Fig. \ref{fig:figure 3} ($a$), is displayed in figure \ref{fig:figure 3} \textit{b}) versus the height of the shell and $\sigma$ ($R_1/R_2$), for different radius of the inner shell~(columns) and different materials~(rows). Consistent with Fig. \ref{fig:figure 3} ($a$), perpendicular anisotropy is promoted by a high aspect ratio and small shell thickness ($R_2-R_1$). In these diagrams, we clearly see that in the case of perpendicular anisotropy, the regions with highest stability are those for rather thick shells~(low $\sigma$), which comes from its large volume and therefore a large energy barrier. Also, as is the case with standard PSA pillars, the stability scales with $M_\mathrm{s}^2$.

We now turn to considering both the core and the shell in dipolar interaction to evaluate how the several geometrical parameters can be best tuned to enhanced stability. In addition to $R_1$, $R_2$ and $L$ already described, we introduce $R_0$ as the radius of the core cylinder, with $R_0<R_1<R_2$.

As described in \cite{BeleggiaShell2009}, the magnetostatic coupling energy of the core and the shell ($E_\textrm{cs}$) is given by the interaction between the shape amplitude of the two objects:
\begin{multline}
    E_\textrm{cs} = \frac{\mu_0 M_{\textrm{core}} M_{\textrm{shell}}}{8\pi^3} \\ \int \frac{\textrm{d}\mathrm{\textbf{k}}}{k^2} D_{\textrm{core}}(\textbf{k})D_{\textrm{shell}}(\textbf{k})(\mathrm{\textbf{m}}_1\cdot \textbf{\textrm{k}})(\textrm{\textbf{m}}_2\cdot \textbf{\textrm{k}}),
    \label{eq:dipolar_shape}
\end{multline}
where $D_{\textrm{core}}$ and $D_{\textrm{shell}}$ are defined by equations \ref{eq:ShapeAmplitude} and \ref{eq:ShapeAmplitude-shell}, respectively. Equation \ref{eq:dipolar_shape} can be formally simplified to:
\begin{multline}
    E_\textrm{cs} = \pi \mu_0 M_{\textrm{core}} M_{\textrm{shell}} R_0 \\ \big[2\cos\theta_1\cos\theta_2 - \sin\theta_1\sin\theta_2\cos(\phi_1 - \phi_2)\big] \\ 
    \times \int^{+ \infty }_{0} \frac{\mathrm{d}k_\perp}{{k_\perp^2}} \bigg[R_2\mathcal{J}_1(k_\perp R_2)\mathcal{J}_1(k_\perp R_0) - \\ R_1\mathcal{J}_1(k_\perp R_1)\mathcal{J}_1(k_\perp R_0)\bigg]\left(1 - e^{-k_\perp L}\right)\;.
    \label{eq:dipolar-main}
\end{multline}
Compared with the initial approach \cite{BeleggiaShell2009, richter_analysis_2006}, we expand the modeling to the case in which the shell is not in contact with the magnetic shell. Making use of $q_i = k _\perp R_i$, $\sigma_i = R_0 / R_i$ and $\tau_i = L / 2R_i$ (where $i$ is related to the inner radius $R_1$ and outer radius $R_2$) the integral of Eq. \ref{eq:dipolar-main} it can be further expanded to 
\begin{multline}
    \mathcal{J}_{cs} = \frac{1}{R_0^2}\int^{+ \infty }_{0} \frac{\mathrm{d}q_2}{\sigma_2^2q_2^2} \mathcal{J}(q_2)\mathcal{J}(\sigma_2q_2)\left(1-e^{-2q_2\tau_2}\right) - \\ \int^{+ \infty }_{0} \frac{\mathrm{d}q_1}{\sigma_1^2q_1^2} \mathcal{J}(q_1)\mathcal{J}(\sigma_1q_1)\left(1-e^{-2q_1\tau_1}\right).
\end{multline}
Making use of the last results from equations  \ref{eq:Shell-demag} and \ref{eq:A11}, it is possible to get the results for the dipolar coupling of core and shell $\mathcal{J}_{cs}$ as 
\begin{multline}
    \mathcal{J}_{cs}R_0^2 = \frac{1}{\sigma_2^2}\left[\sigma_2F_{0}(\sigma_2^2) + A_{1,1}^{-1}(2\tau_2, 1, \sigma_2)\right] - \\  \frac{1}{\sigma_1^2}\left[\sigma_1F_{0}(\sigma_1^2) + A_{1,1}^{-1}(2\tau_1, 1, \sigma_1)\right].   
\end{multline}
In addition, these results can be linked to the demagnetising factors of the shell and core through equation \ref{eq:Shell-demag}.

\begin{figure}
    \centering
    \includegraphics{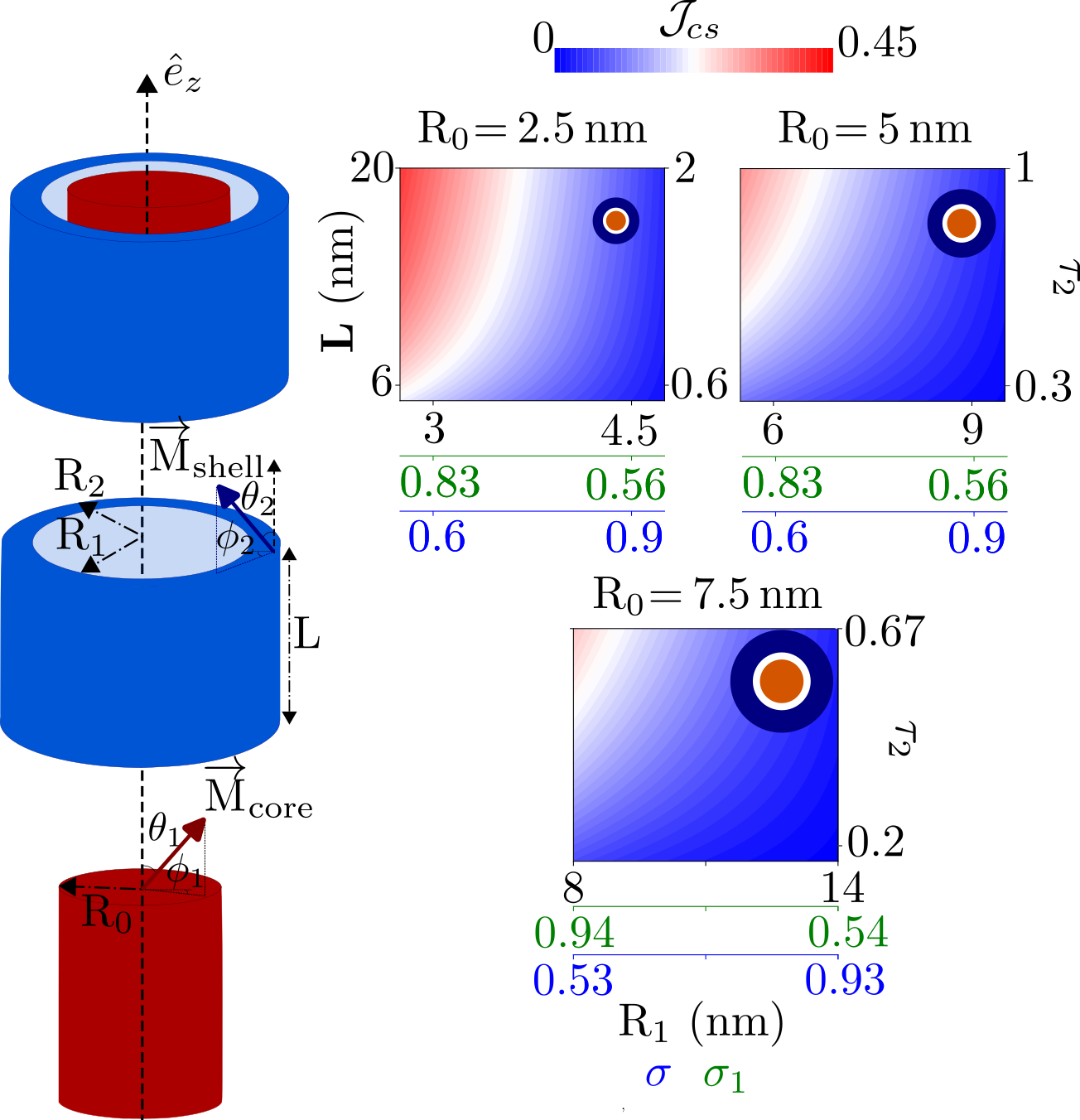}
    \caption{(left side) Schematics of the core-shell system where $R_0$ is the inner radius of the magnetic core (red colour) with magnetisation orientation $\overrightarrow{M}_\mathrm{core}$ with angles $\theta_1, \phi_1$; magnetic shell (blue color) with magnetisation orientation $\overrightarrow{M}_\mathrm{shell}$ with angles $\theta_2, \phi_2$. Both parts have a common thickness of L. Right side: dipolar coupling magnitude $\mathcal{J}_\mathrm{cs}$ for a fixed value of $\sigma_2 = 0.5$ for different $R_0$, as a function of its thickness L and inner radius $R_1$.} 
    \label{fig:figure 5}
\end{figure}

In the end, the magnetostatic coupling energy depends on the geometry, and it is proportional to the magnetisation of both the shell and the core:

\begin{multline}
    E_\textrm{cs} = \mu_0\pi R_0^3M_\mathrm{core}M_\textrm{shell} \times \underbrace{\mathcal{J}_{cs}(\tau_0, \tau_1, \tau_2, \sigma_1, \sigma_2)}_{\textrm{Geometrical scaling factor}} \\ \times (2\cos\theta_1\cos\theta_2 - \sin\theta_1\sin\theta_2).
    \label{eq: Scaling Dipolar}
\end{multline}

The dependency of the scaling factor ($\mathcal{J}_{cs}$) versus the system's geometry is illustrated in Fig. \ref{fig:figure 5} for different values of core radius $R_0$ and keeping constant the ratio $\sigma_2 = 0.5$. Thus, the three plots differ in core radius, separation between the inner shell and core, as well as the ratio between the inner and outer radius of the shell $\sigma_s = R_1 / R_2$. It is observed that a stronger geometrical scaling is achieved at thinner separations and longer aspect-ratios. Besides tuning $\mathcal{J}_{cs}$, The total magnetostatic coupling energy can be further increased through an increase in $R_0$ , related to the magnetic volume, or through the increase in the magnetisation of both shell and magnetic core $M_\textrm{core}M_\textrm{shell}$. 

As a final step, we now compute the total energy of the dipolar-coupled core-shell magnetic system $E_\mathrm{tot}$ as \cite{BeleggiaShell2009}
\begin{multline}
    E_\mathrm{tot}(\theta_1, \theta_2,\phi_1,\phi_2) = E_\mathrm{c}\sin^2{\theta_1} + E_\mathrm{s}\sin^2{\theta_2} + \\ E_\textrm{cs} (\theta_1, \theta_2,\phi_1,\phi_2)\;,
    \label{eq:EB-system}
\end{multline}
where $E_\mathrm{c}$ is the energy barrier of the isolated magnetic core, $E_\mathrm{s}$ is that of the  shell alone, and $E_\textrm{cs}$ the coupling energy. For the sake of simplicity and generality, we write the energy of the system using the coefficients $\mathcal{A}$, $\mathcal{B}$, $\mathcal{C}$ and $\mathcal{D}$ for the different angular dependencies: 
\begin{multline}
    E_\mathrm{tot} =\mathcal{A}\sin^2{\theta_1} +\mathcal{B}\sin^2{\theta_2} - \\ \mathcal{C}\sin\theta_1\sin\theta_2\cos(\phi_1 - \phi_2) + \mathcal{D}\cos\theta_1\cos\theta_2\;.
    \label{eq:EB-system-simplified}
\end{multline}

$\mathcal{A}$ is the self energy barrier of the magnetic core, $\mathcal{B}$ the self energy barrier of the magnetic shell, $\mathcal{C}$ and $\mathcal{D}$ are the core-shell coupling energies in the parallel state for the transverse and axial directions, respectively. With the present geometry all four coefficients are positive and $\mathcal{D} = 2C$, see Eq. \ref{eq:dipolar-main}. Thus, the lowest-energy state is expected when both shell and core are out-of-plane-magnetised and antiparallel on to one another, and the most energetic state when both shell and core are in-plane-magnetised and antiparallel to one another. More generally, for arbitrary $\theta_1$ and $\theta_2$ the inspection of Eq. \ref{eq:EB-system} shows that the minimum energy path for switching is expected to satisfy $\phi_1=\phi_2 + \pi$ at all time, which we consider satisfied for now on. 

\begin{figure}[h!]
    \centering
    \includegraphics{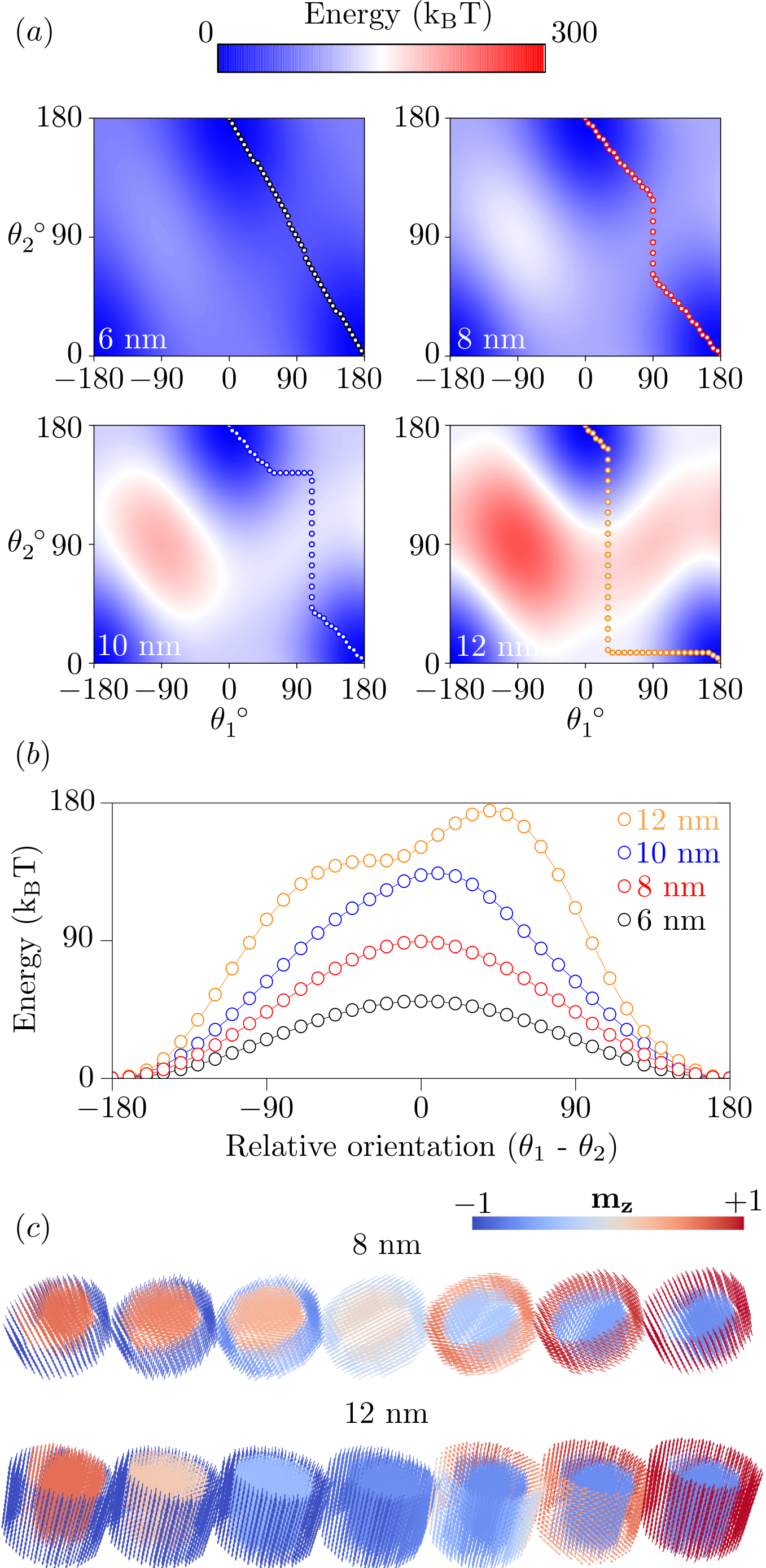}
    \caption{(\textit{a}) Angular dependency of the energy barrier of the magnetic system for different core-shell length with a fixed core radius of $\SI{7}{nm}$, inner shell radius of $\SI{8}{nm}$ and outer radius of $\SI{10}{nm}$, for the saturation magnetisation of $\SI{1}{MA/m}$ for the magnetic core and $\SI{1.446}{MA/m}$ for the magnetic shell. Negative angles can be viewed as standing for the case $\phi_1=\phi_2+\pi$, while formally only positive angles are defined in spherical coordinates. The trace of the minimum energy path for each geometry is shown with filled white circles. (\textit{b}) Relative orientation of the minimum energy path for increasing thickness of the core-shell system and (\textit{c}) 3D snapshots for the height of 8 and $\SI{12}{nm}$, along the minimum energy path.}
    \label{fig:figure 6}
\end{figure}

The angular dependence of the energy barrier described by equation \ref{eq:EB-system-simplified} is shown in Fig. \ref{fig:figure 6} $(a)$, for the case of a fixed core radius of $\SI{7}{nm}$, a shell inner radius of $\SI{8}{nm}$ and an outer radius of $\SI{10}{nm}$ for several lengths (6, 8, 10 and $\SI{12}{nm}$). The core magnetisation is $\SI{1}{MA/m}$, the surface anisotropy is $\SI{1.4}{mJ/m^2}$ and the shell magnetisation is $\SI{1.446}{MA/m}$. The corresponding coefficients $\mathcal{A}$, $\mathcal{B}$ and $\mathcal{C}$ involved in the dipolar energy are listed in Tab.\ref{tab:energyCoefficients}.  For each vertical aspect ratio the minimum energy path is computed \cite{marcos-alcalde_mepsa_2015} and displayed in Fig. \ref{fig:figure 6} (\textit{b}).  

\begin{table}[h!]
\caption{Coefficients of the energy of the coupled core-shell systems shown on Fig. \ref{fig:figure 6} in units of k$_\textrm{B}$T, i.e., for different heights and fixed core-shell radius of $R_0 = \SI{7}{\nano\meter}$, $R_1 = \SI{8}{\nano\meter}$ and $R_2 = \SI{10}{\nano\meter}$.}
\centering
\label{tab:energyCoefficients}
\begin{tabular}{cccc}
\hline\hline
Height (nm)& $\mathcal{A}$ ($k_\textrm{B}T$)& $\mathcal{B}$ ($k_\textrm{B}T$)& $\mathcal{C}$ ($k_\textrm{B}T$)\\
\hline
6& 15 & 17  & 19 \\
8& 22 & 43  & 25 \\
10& 30 & 72  & 32 \\
12& 42 & 100 & 35\\
\hline\hline
\end{tabular}
\end{table}

In all four cases, we confirm that the ground state is perpendicular anisotropy in an antiparallel state. The energy barrier increases with the vertical aspect ratio, as expected from our previous discussion on individual core, individual shell, and magnetostatic coupling. For the system with height $\SI{6}{nm}$ the minimum energy path is a line with slope $-1$, which means that the two parts rotate simultaneously. We indeed expect this if the coupling between shell and core is comparable or even larger than the anisotropy energy of each element taken separately, which is the case here. To the contrary, in the system with height $\SI{12}{nm}$ the coupling is smaller than any of the energy barriers of the two elements taken independently. This implies that a sequential reversal allows to minimise the total energy barrier, which indeed is seen in the energy maps and the energy plot in Fig.\ref{fig:figure 6} (b). The heights $\SI{8}{nm}$ and $\SI{10}{nm}$ appear as cross-over situations, with a combination of sequential and coupled paths. From a device point of view, a strong coupling situation may imply a prohibitively-large energy barrier, while the low-coupling one is not suited either if only the core is connected through the leads for STT switching, as the coupling is not sufficient to switch the shell once the core has been switched. Thus, we expect that the optimum situation is to be found in the cross-over regime, to be identified using simulation.  

The discussion of the energetics of a coupled set of macrospins and its minimum-energy path is valuable to draw the general trends and identify the suitable range of operating parameters. Now, we turn to micromagnetic simulations to refine the conclusions and determine the optimal geometrical and magnetic parameters in the framework of the PSA-MTJ, taking into account precessional effects, the excitation means~(current injected via an MTJ in the core only) and possible non-macrospin effects. For this purpose, we use a finite-difference 3D micromagnetic solver with a cubic computational cell of $\SI{1}{nm^3}$ \cite{Nuno, buda_micromagnetic_2002}, in which the magnetisation dynamics is described by the Landau-Lifshitz-Gilbert-Slonzcewski (LLGS) equation. Here we consider that an injected spin-polarised current interacts with the magnetic body of the core via a damping-like torque ($\Gamma_{\textrm{STT}}$), while we neglect the effect of the field-like torque:
\begin{equation}
    \partial_t\textbf{m} = -|\gamma|\mu_0(\textbf{m}\times \textrm{H}_{\textrm{eff}}) + \alpha(\textbf{m}\times\partial_t\textbf{m}) + \Gamma_{\textrm{STT}}\;,
\end{equation}
with $\gamma$ the gyromagnetic ratio, and: 
\begin{equation}
    \Gamma_{\textrm{STT}} = -|\gamma|a_\parallel \textrm{V} \textbf{m}\times(\textbf{m}\times\textbf{m}_{\textbf{RL}}),
\end{equation}
where V is the applied bias voltage, \textbf{m} is the unit vector for magnetisation of the storage layer at the location considered, \textbf{m}$_{\textrm{RL}}$ is the unit vector for magnetisation of the reference layer and $a_\parallel$ is the STT pre-factor. The latter is expected to physically decay along the thickness of the magnetic body through the interaction with magnetisation \cite{slonczewski_theory_2007, sun_magnetoresistance_2008}. Contrary to the case of ultrathin MTJs, the torque is expected to be largely nonhomogeneous and predominantly exerted close to the interface with the tunnel barrier.  To reflect this, we apply the full STT torque in the first layer of cells, whose strength matches the expected torque integrated when moving away from the interface, with an exponential decay 
\begin{equation}
    \langle a_\parallel \rangle = a_\parallel(0) \times \sum_{n=0}^{N-1} \exp{\frac{-n\lambda_{STT}}{\delta_z}}\;,
\end{equation}
in which $\lambda_{\textrm{STT}}$ is the characteristic STT decay length (taken as $\SI{1}{nm}$) and $\delta_z$ is the cell size ($\SI{1}{nm}$). Making use of the definition 
\begin{equation}
a_\parallel(0) = \frac{\hbar}{2e}\frac{\eta_{\textrm{STT}}}{\textrm{R}\times\textrm{A}}\frac{1}{M_\mathrm{s}\delta_z}
\end{equation}
where $\eta_{\textrm{STT}}$ is the STT efficiency \cite{sun_spin-torque_2013}, it is possible to calculate the strength of the equivalent interfacial STT pre-factor. In this study, we consider that the core and thus the interface with Mg0 is made of a FeCoB base alloy, with spontaneous magnetisation of $M_\mathrm{s} = \SI{1}{MA/m}$, exchange stiffness of $A_\mathrm{ex} = \SI{15}{pJ/m}$, and a damping value of $\alpha = 0.01$, $a_\parallel=\SI{140}{mT/V}$ considering a typical TMR value of 100$\%$ and a R$\times$A product of $\SI{1.5}{\Omega\cdot\mu m^2}$. The voltage is only applied at the interface of the the magnetic core, so that the dynamics of the magnetic shell may only result from its dipolar interaction with the core. Surface anisotropy is considered to only be present at the interface between the core and the tunnel barrier, with a value of $\SI{1.4}{mJ/m^2}$.

\begin{figure}
    \centering
    \includegraphics{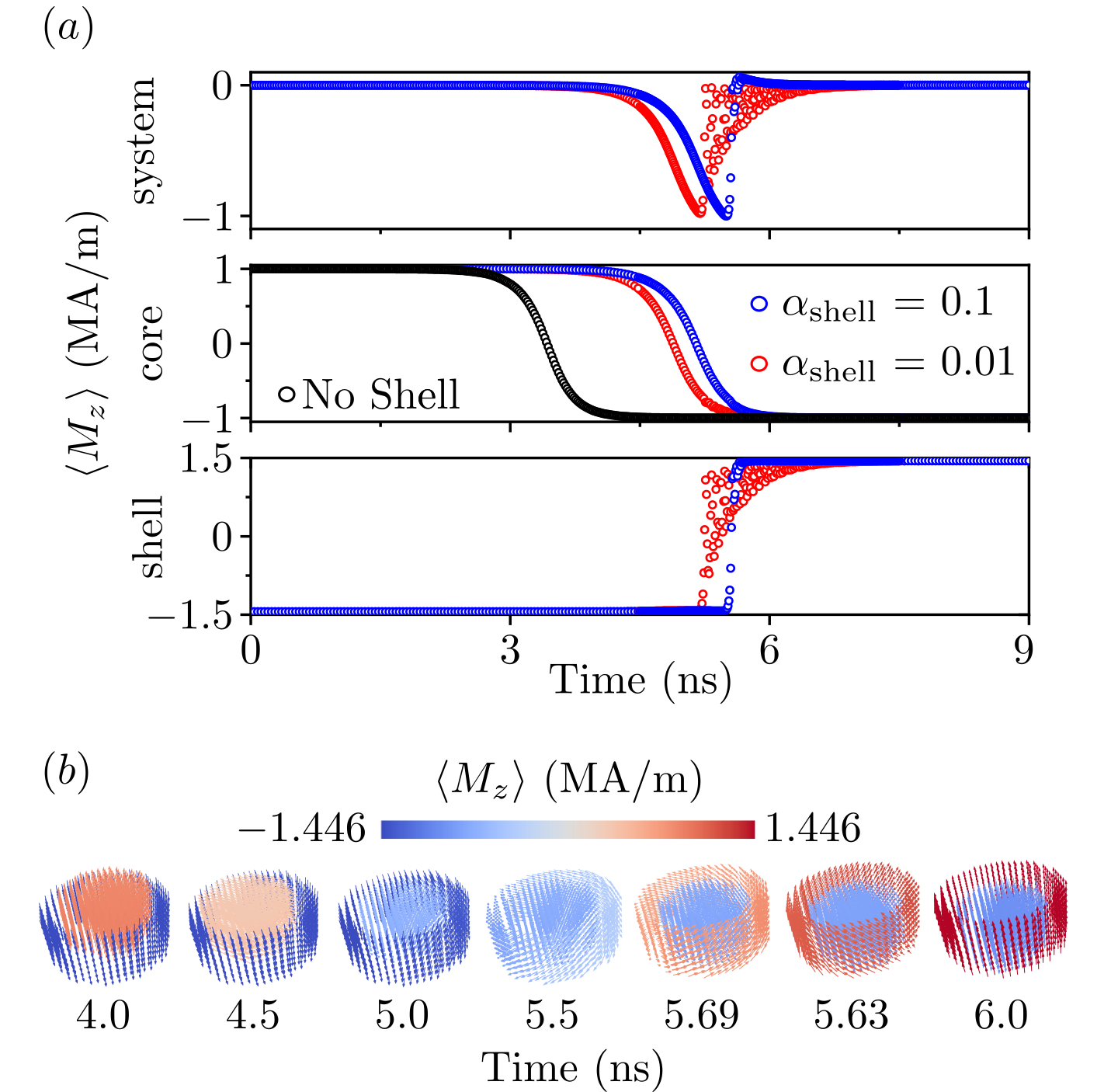}
    \caption{($a$) Time trace of the mean magnetisation components $\langle M_z \rangle$ for a core-shell system, considering two different damping parameters for a Co magnetic shell. The top plots shows magnetisation averaged over the entire composite system, while the bottom plots show the components of the core and shell independently. ($b$) 3D snapshots of the time evolution of a core-shell system under an applied voltage of $\SI{-1}{V}$ for the situation of low and high damping. Colorbar shows the magnitude of the z component of the average magnetisation in each computational cell.}
    \label{fig:figure 7}
\end{figure}

We considered the switching dynamics for two different material parameters for the magnetic shell. The first situation makes use of the system presented in Fig. \ref{fig:figure 6} with a length of $\SI{8}{nm}$ (chosen such that the stability determined from the minimum energy path is around $\SI{80}{k_BT}$), $R_0 = \SI{7}{nm}$, $R_1 = \SI{8}{nm}$ and $R_2 = \SI{10}{nm}$, where a magnetic shell of Co is used, with $M_\mathrm{s}$ of $\SI{1.446}{MA/m}$, exchange stiffness $A_\mathrm{ex}$ of $\SI{15}{pJ/m}$ and damping either $\alpha = 0.01$ or $\alpha = 0.1$. The applied voltage is $\SI{-1}{V}$.  The time trace of the mean perpendicular magnetisation components of the core, shell, and full system are shown in Fig. \ref{fig:figure 7}. The magnetisation switching of the isolated magnetic core (without magnetic shell) happens at an earlier stage compared with the core coupled with a magnetic shell. This is expected, since the core alone has a lower stability, around $\SI{22}{k_BT}$, and it is also not stabilised along the initial direction by the stray field of the shell. When considering the shell, we can see that its reversal is delayed with that of the core, which makes sense, as current is injected only into the latter. Besides, the switching of the core is delayed in the case of a shell with higher damping, which we attribute to an incubation effect, with a lower efficiency to set in precessional dynamics: part of the injected energy may be absorbed in the shell, via dipolar coupling. A striking effect is that for lower damping of the shell, even though both the core and the shell start their reversal faster, the shell takes longer to relax to the stable final state. Indeed, if the damping of the shell is large, then the incubation time for precession is longer, but then there is a faster relaxation to the stable state. This is consistent with the large oscillations of perpendicular magnetisation seen in the case of lower damping.  

\begin{figure}[h]
    \centering
    \includegraphics{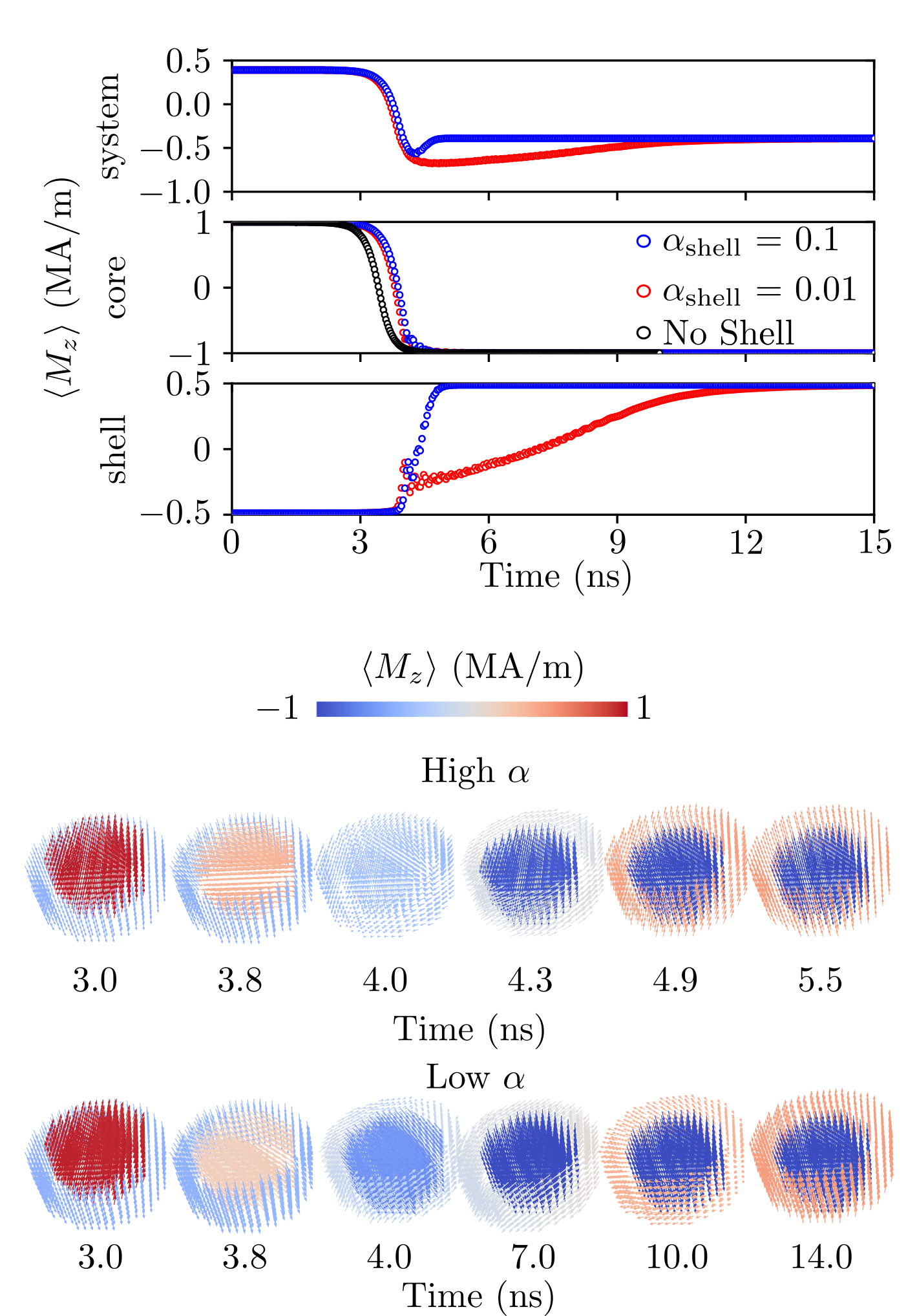}
    \caption{($a$) Time trace of the mean magnetisation components $\langle M_z \rangle$ for a core-shell system, considering two different damping parameters for a Ni magnetic shell. The top plots shows magnetisation averaged over the entire composite system, while the bottom plots show the components of the core and shell independently. ($b$) 3D snapshots of the time evolution of a core-shell system under an applied voltage of $\SI{-1}{V}$ for the situation of low and high damping. Colorbar shows the magnitude of the z component of the average magnetisation in each computational cell.}
    \label{fig:figure 8}
\end{figure}

We now consider a shell made of a low-magnetisation material, considering Ni with spontaneous magnetisation of $\SI{489}{kA/m}$ and exchange stiffness of $\SI{8}{pJ/m}$, again for both low and high damping situations~(Fig. \ref{fig:figure 8}). As concerns the switching time of the core, it is seen that it is similar to the one without a magnetic shell, both in incubation time and switching dynamics, and it is also barely dependent on the damping parameter of the shell. This is understandable as the lower stray field off the shell has a lower stabilising effect than that shown on Fig. \ref{fig:figure 7}; and also it will be less effective in pumping energy out of the initial oscillations during the incubation time. As concerns the shell, a close examination of the time scale shows that its reversal starts when the core has not fully reversed, but rather has an in-plane magnetisation. Besides, its reversal takes longer than for a high-magnetisation shell~(figure \ref{fig:figure 7}), especially for low damping. This may result from the small energy barrier of the magnetic shell (around $\SI{5}{k_BT}$), so that it is mostly and directly affected by the dipolar coupling with the magnetic core. Figure \ref{fig:figure 8} (\textit{b}) shows 3D snapshots of the reversal for the high and low damping situations, illustrating the ringing effect in the latter.

In conclusion, all geometrical and magnetic parameters influence the switching process in a core-shell process. Unlike the case of simple MRAM cells, either flat or PSA, higher damping may be beneficial in the shell. In practice, the parameter space shall probably be explored in more detail to meet the desirable criteria of any given situation, in terms of stability, switching time and power consumption. 

\section{Reduced cross-talk making use of the core-shell concept}
\label{sec-crossTalk}

We examine here another advantage of the core-shell storage layer, which is the possibility to mitigate the mutual coupling via stray field from neighboring bits in a high dense array via the optimisation of both geometry and magnetic properties. To evaluate the gain quantitatively, we compute the magnetic field emanating from a pillar with the formalism implemented in \cite{Taniguchi}. Starting from the magnetic potential, it is possible to derive the radial ($H_r$) and perpendicular components ($H_z$) of the magnetic field, respectively:

\begin{multline}
H_r(r,z) = 2\pi R_0 M_\mathrm{s} \int_{0}^{\infty} \mathrm{d}k \big[e^{-k|Z-L|} - e^{-k|Z|}\big] \\ \mathcal{J}_{1}(kr)\mathcal{J}_{1}(kR_0).
\end{multline}

\begin{multline}
H_z(r,z) = 2\pi R_0 M_\mathrm{s} \int_{0}^{\infty} \mathrm{d}k \big[\mathrm{sign}(Z - L)e^{-k|Z-L|} \\ - \mathrm{sign}(Z)e^{-k|Z|}\big] \mathcal{J}_{0}(kr)\mathcal{J}_{1}(kR_0)\;,
\end{multline}
where $\mathcal{J}_x$ are the Bessel functions of the order $x$, $R_0$ is the radius of the magnetic cylinder, $L$ its height and $M_\mathrm{s}$ its spontaneous magnetisation. These equations can be used to calculate both the demagnetising field (inside the magnetic body $0 < Z < L$ and $r<R_0$, since there is a discontinuity for Z = L and Z = 0 and $r < |R_0|$) and the stray field outside (for $Z > L$ and $r > |R_0|$). 

\begin{figure}
    \centering
    \includegraphics{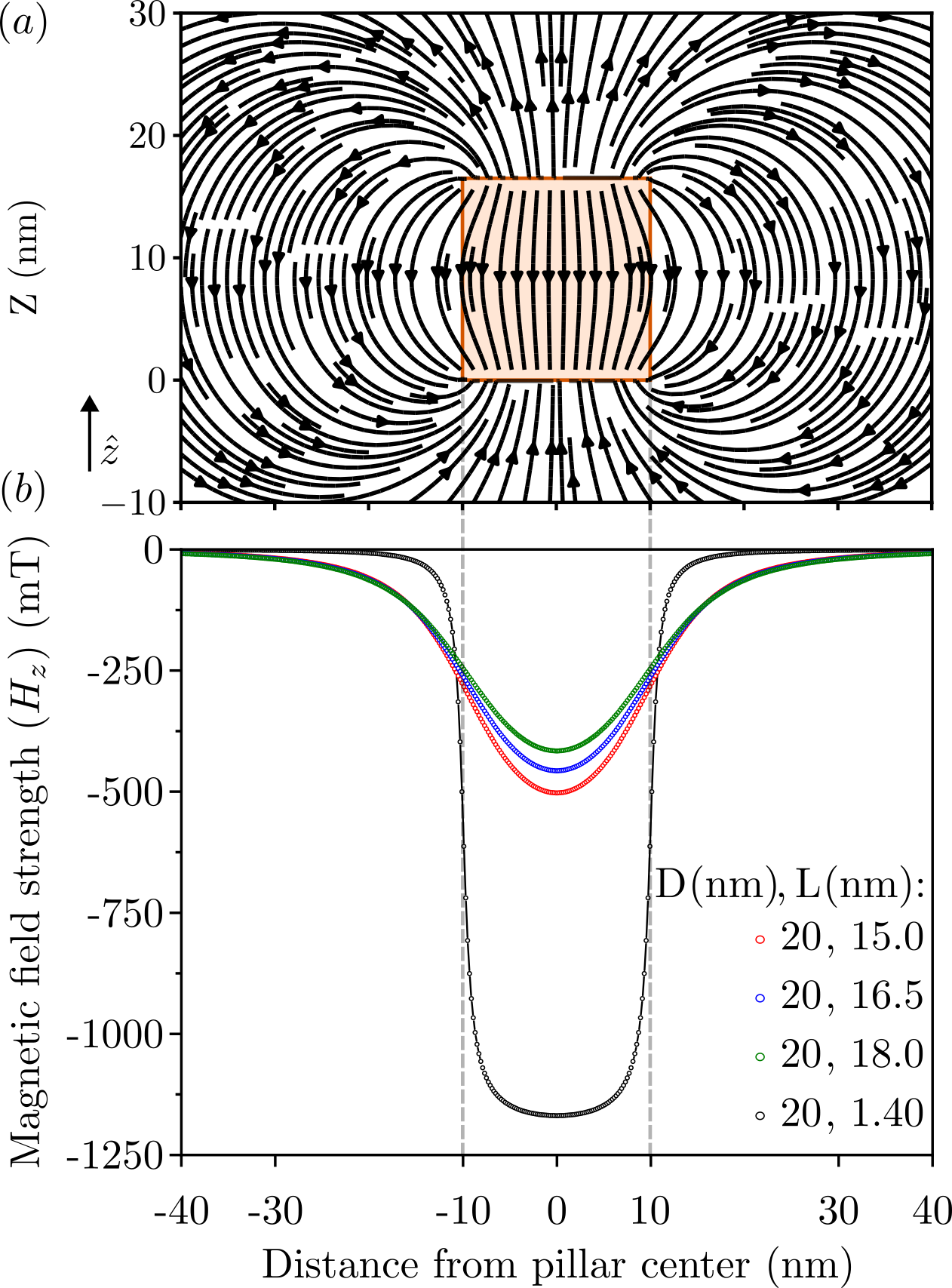}
    \caption{(\textit{a}) 2D stray field lines arising from a pillar with high aspect ratio (thickness of $\SI{16.5}{nm}$ and $\SI{20}{nm}$ diameter, orange-outlined) with up magnetisation along $\hat{z}$. (\textit{b}) magnetic field strength at mid thickness of the magnetic pillar, for 4 different aspect ratios.}
    \label{fig:figure 9}
\end{figure}

hese are displayed in Fig. \ref{fig:figure 9} (\textit{a}) for a magnetic pillar with $2R_0=\SI{20}{nm}$ diameter and height $L=\SI{16.5}{nm}$. We consider perpendicular magnetisation, which in practice would arise from the combination of shape and surface anisotropy. The magnetic field strength at midheight of the pillar is shown in Fig. \ref{fig:figure 9} (\textit{b}) for pillars with different aspect ratios, all assumed to have perpendicular magnetisation. In the situation of a low aspect ratio (such as the usual p-MTJ with \SI{20}{nm} diameter and \SI{1.4}{nm} height), we can observe a thin-film-like demagnetising field for $r < |R_0|$, with a sharp decay outside the pillar. For PSA-like pillars the demagnetising field is reduced, but the stray field is greatly enhanced due to the larger volume involved and hence the magnetic moment. As in two dimensions, the extension of stray field is short-ranged, typically of the order of the system thickness or so \cite{fruchart_high_1998}, one may evaluate its impact based on a finite and small array, which we do in the following.

\begin{figure}
    \centering
    \includegraphics{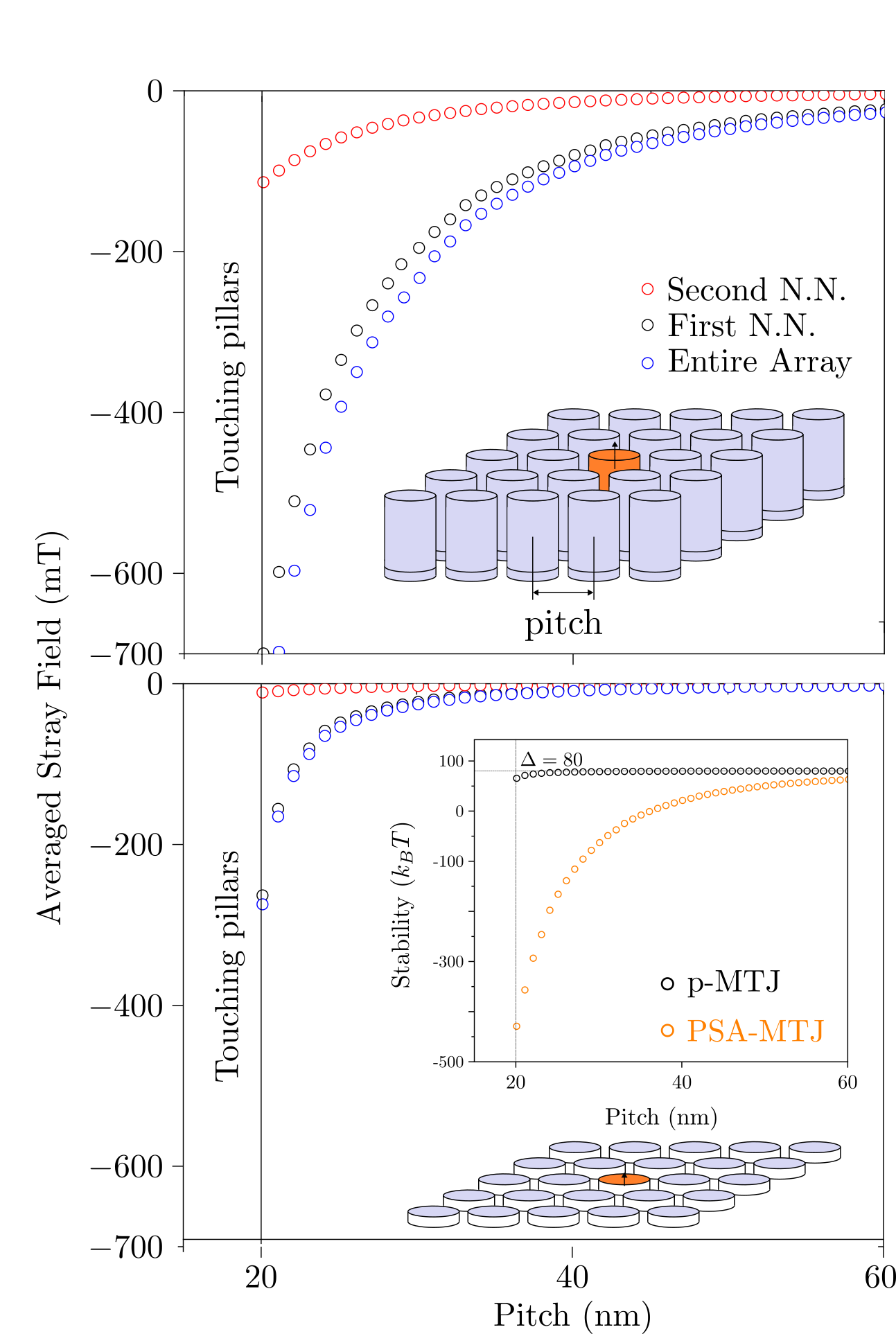}
    \caption{Volume-averaged stray field at the central pillar in a $5\times5$ array for (top) PSA-MTJs with a diameter of $\SI{20}{nm}$ and height $\SI{16.5}{nm}$ and (bottom) p-MTJs with a diameter of $\SI{20}{nm}$ and thickness $\SI{1.4}{nm}$. The contribution from the first shell of neighbours is shown with black open disks. The contribution of the second shell of neighbours is shown with red open disk. The sum of both contributions is shown with open blue disks. The dependency of the thermal stability as a function of the pitch for both PSA-MTJ and p-MTJ is shown as an inset.}
    \label{fig:figure 10}
\end{figure}

To evaluate the strength of stray field in a dense array of pillars with perpendicular magnetisation, we consider a 5x5 pillar array, as depicted in the inset of the Fig. \ref{fig:figure 10}. We define the pitch as the distance between the center of two adjacent pillars, and thus, when in contact, it equals the pillar diameter. We do not consider the stray field arising from a possible reference layer below the pillar, as this can be engineered to be negligible \cite{veiga_control_2023}. 
\begin{figure}
    \centering
    \includegraphics{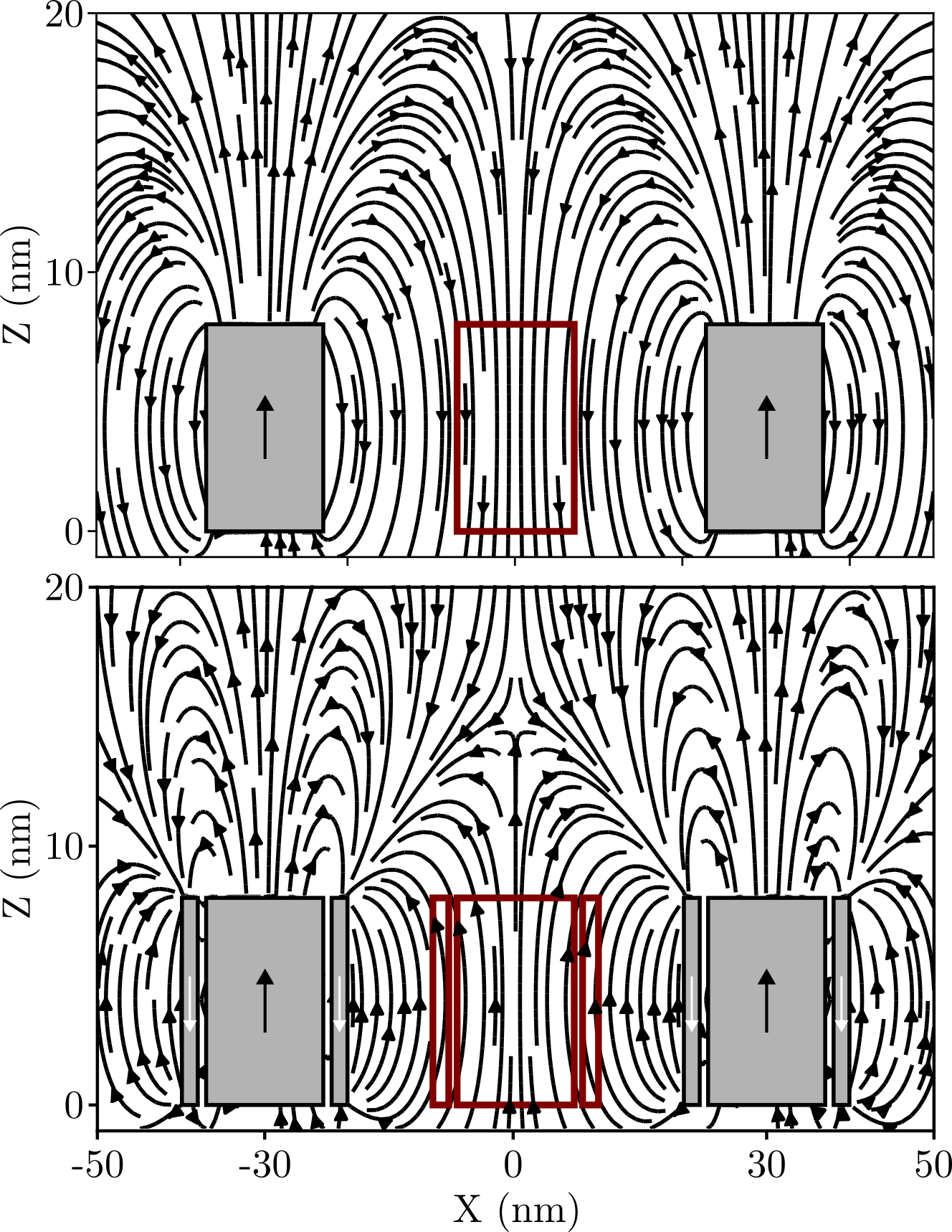}
    \caption{2D stray field lines of (\textit{a}) two magnetic cores with a pitch of $\SI{30}{nm}$. Each magnetic core has a diameter of $\SI{14}{nm}$ and a height of $\SI{8}{nm}$ and (\textit{b}) two core-shell systems with a pitch of $\SI{30}{nm}$. Each system is composed of a magnetic core with a radius of $\SI{7}{nm}$, inner radius $\SI{8}{nm}$, an outer radius $\SI{10}{nm}$, and a height of $\SI{8}{nm}$. In both situations, the system is shown with a filled gray color, and the magnetisation direction of the core is shown in black, pointing up. For the core-shell situation, the magnetisation of the shell is white, pointing down. In both cases, a central element is shown with an orange colour.}
    \label{fig:figure 11}
\end{figure}

Figure \ref{fig:figure 10} shows the stray field in an array as a function of its pitch, for pillars of high vertical aspect-ratio ($\SI{20}{nm}$ diameter and $\SI{16.5}{nm}$ height, upper panel), and for cells with a flat aspect ratio ($\SI{20}{nm}$ diameter and $\SI{1.4}{nm}$ height, lower panel), both with spontaneous magnetisation of $\SI{1}{MA/m}$. These situations correspond, respectively, to PSA-MTJs and p-MTJs. Both panels show the contribution of the first shell, the second shell, and the total. For p-MTJs, the stray field is significant when pillars are in contact, but becomes negligible for pitch $\SI{50}{nm}$ or greater, as is standard for usual integration processes \cite{wan_fabrication_2022}. The impact of the second shell of neighbors is insignificant in all cases, so considering the impact of the first nearest neighbors is enough \cite{wu_impact_2020}. For PSA MTJs, the stray field from the first shell is stronger, and that of the second is non-negligible. The next shells may be safely ignored thanks to the $1/r^2$ decay of the stray field emanating from a complete shell. The resulting total stray field is very significant and is expected to affect the stability factor of the device. It could be increased (reduced) if the average stray field points in the same (opposite) direction as the pillar magnetisation  \cite{khvalkovskiy_basic_2013}: 
\begin{equation}
    \Delta_{\mathrm{k}_\mathrm{B}\mathrm{T}}^{\textrm{stray}} = \Delta_{\mathrm{k}_\mathrm{B}\mathrm{T}}\left(1 \pm \frac{H_z^{\textrm{stray}}}{H_\textrm{K}^{\textrm{eff}}}\right)\;,
\end{equation}

where $H_\textrm{K}$ is the effective anisotropy field of the magnetic element and $H_z^{\textrm{stray}}$ the average stray field felt by it. Both increase and decrease of stability are detrimental for device operation, the former related to the write operation, the latter related to the thermal stability \cite{wu_impact_2020}. The dependence of the stability with the pitch of the array is shown in the inset of Fig. \ref{fig:figure 10} (bottom panel). In the case of an array of p-MTJs it is seen that a pitch around $\SI{30}{nm}$ is sufficient to avoid a significant variation in stability. This is related to the fact that in 2D systems the range of dipolar interactions scales with the height of the system.  To the contrary, for PSA-MTJ, it would be necessary to drastically increase the pitch to $\SI{100}{nm}$ or more to reach the same level of crosstalk reduction. The stray field would even be enough to flip magnetisation of the central pillar smaller than $\SI{50}{nm}$.  

\begin{figure}[h!]
    \centering
    \includegraphics{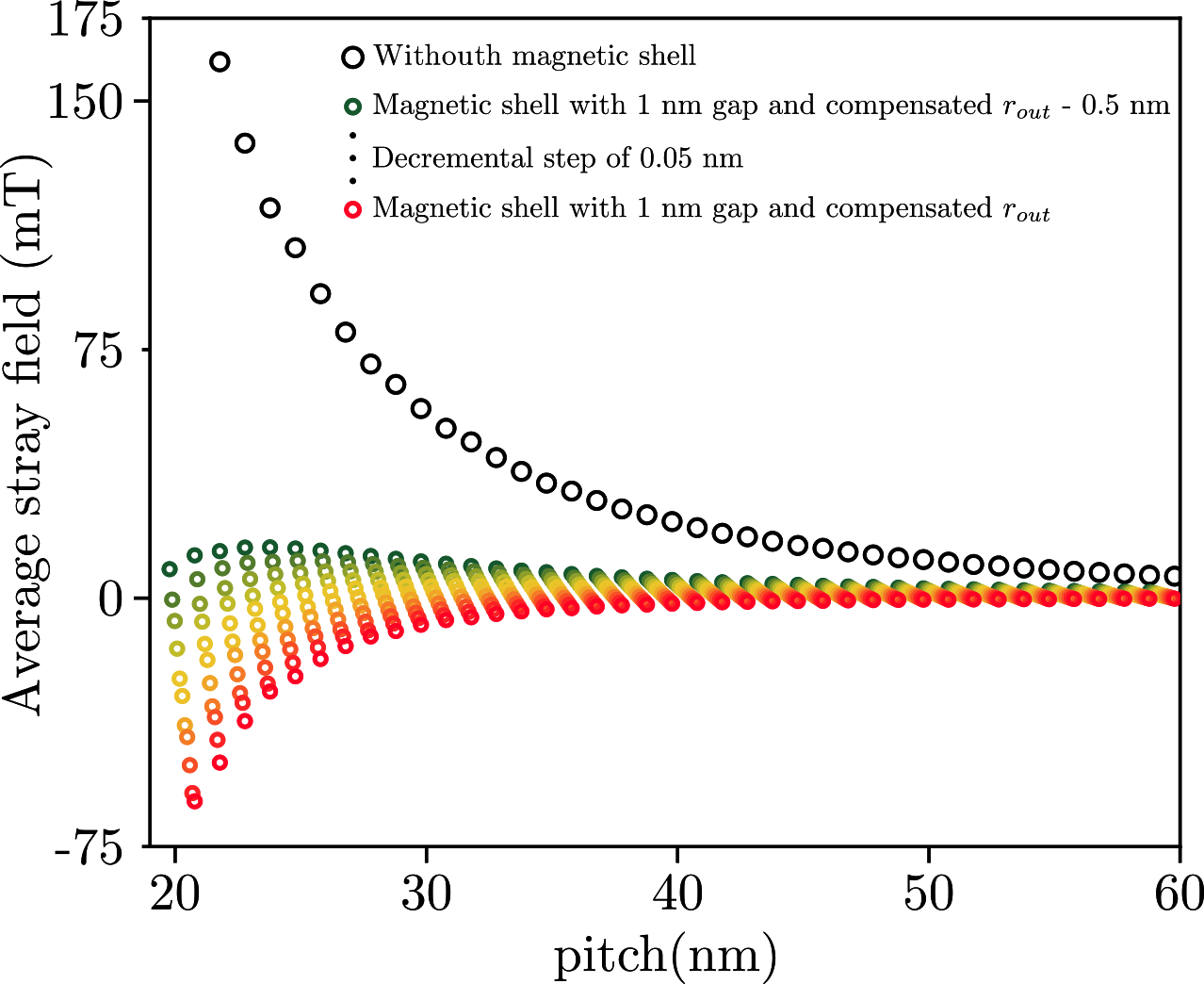}
    \caption{Volume-averaged stray field at the central pillar in an array of PSA-MTJs with a diameter of $\SI{20}{nm}$ and height $\SI{16.5}{nm}$ as a function of the pitch for situation of unshelled pillar (open black circles) and shelled pillars (colored open circles) for an inner radius of $\SI{8}{nm}$ and different outer radius, from compensated magnetic volume (red open circles) to a $\SI{0.5}{nm}$ smaller outer radius. The system is shown in the inset where the neighbouring core-shell have the opposite orientation of that of the central core.}
    \label{fig:figure 12}
\end{figure}

We expect that the core-shell design can also contributs to a mitigation of the stray field from the neighboring dots, thanks to the antiparallel and mostly compensated core and shell magnetic moments. This is illustrated in Fig. \ref{fig:figure 11} (\textit{b}), considering the effect of two neighboring cells with center-to-center distance of $\SI{30}{nm}$, each with a CoFeB core with $\SI{14}{nm}$ diameter and $\SI{8}{nm}$ height. The top panel shows the case of the simple pillar, while in the bottom panel these are surrounded by a shell of Co with an outer radius of $\SI{10}{nm}$ The field lines are indeed drastically different in the core-shell system, characteristic of a quadrupolar arrangement, as the core and shell moments are nearly balanced, in contrast with the situation with no magnetic shell, characteristic of a dipolar system. 

Figure \ref{fig:figure 12} quantitatively shows the volume-averaged dipolar field felt by the core of the central cell in a $5\times5$ array, for the case of a core radius of $\SI{7}{nm}$ with spontaneous magnetisation of $\SI{1}{MA/m}$, an inner shell radius of $\SI{8}{nm}$ and a varying outer shell radius from the situation where there is compensation of magnetic volume ($M_\mathrm{c} V_\mathrm{c} = M_\mathrm{s} V_\mathrm{s}$, where $V_{\mathrm{c},\mathrm{s}}$ is the volume from the core or shell) down to smaller outer radius, with a spontaneous magnetisation of $\SI{1.446}{MA/m}$. Small variations in the outer radius lead to qualitative variations of the stray field, which, however, always decays much faster with a magnetic shell than without it~(open black disks). The difference is related to the $\sim1/r^4$ decay of the stray field for quadrupoles.

\section{Conclusion} 

We quantitatively evaluated the advantages of using composite storage cells for perpendicular-shape anisotropy MTJs, made of a core pillar surrounded by a magnetic shell decoupled from exchange but coupled by dipolar interaction, so that it behaves as a synthetic ferrimagnet with out-of-plane magnetisation. The practical realisation of such cells could make use of conformal coating techniques, such as atomic layer deposition for the formation of the magnetic shell, around a patterned magnetic tunnel junction. Using both analytical models and micromagnetic simulations, we outline the benefits in terms of enhanced stability and stray-field mitigation in dense arrays. Due to the antiparallel state further promoting perpendicular magnetisation besides the shape effect of the individual core, it is possible to consider reducing the core-shell height for a given stability factor, thus reducing the switching voltage necessary to trigger reversal and lowering its switching time. Making use of an appropriate geometry for stability of around $\SI{80}{k_BT}$, the use of magnetic shells of Ni versus Co are compared. For both materials, a larger damping allows for a faster reversal, a criteria opposite to the one to follow for classical MTJs . The use of a magnetic shell also dramatic reduces the stray field arising from neighbouring cells, allowing to extend the technology to very dense arrays. These assets are general to core-shell structures, so they remain valid for other geometries, such as prismatic.  

\begin{acknowledgments}
This work was supported by Samsung Electronics Co., LTD. (IO190709-06540-02).
\end{acknowledgments}

\bibliography{references.bib}  

\end{document}